# Infrared Single-Photon Detector based on Silicon Two-Photon Absorption[1]

**Alex Hayat, Pavel Ginzburg and Meir Orenstein**

Department of Electrical Engineering, Technion, Haifa 32000, Israel

ahayat@tx.technion.ac.il

We propose a scheme for infrared single-photon detection based on two-photon absorption at room-temperature in Si avalanche photodiodes, where the detected photon's energy is lower than the bandgap and the energy difference is complemented by a pump field. A quantum non-perturbative model is developed for non-degenerate two-photon absorption in direct and indirect semiconductors yielding proper non-divergent rates allowing device efficiency optimization. The proposed monolithic detector is simple, miniature, integrable and does not require phase matching, while not compromising the performance and exhibiting even better efficiency than the competing up-conversion schemes (~1 order of magnitude) for similar optical pump levels.

---

[1] To appear in Physical Review B (APS copyright) - http://prb.aps.org/

## I. INTRODUCTION

Quantum communications over optical fiber networks employ infrared photons at the wavelength range of 1.3-1.6μm. For these applications, the most widely used single-photon detectors are the narrow-banddgap InGaAs/InP avalanche photodiodes (APDs) with relatively low efficiency and high dark count rates (DC) ($> 10^5$ s$^{-1}$) [1], which requires operating these detectors in gated mode synchronized with the expected photon arrival. Additional important drawback of this detector family is the high afterpulsing probability, which causes long recovery times between detection events, thus limiting the qubit-rate to sub Mqubits/sec. Such low rates present a major obstacle for the incorporation of analytically secure quantum cryptography into the existing fiber-optical communications infrastructure [2]. Silicon APDs have higher quantum efficiency (~70%), much lower DC rate (<100sec$^{-1}$) due to the wider bandgap, do not require gating and have much less severe afterpulsing; however they cannot be applied directly to the telecom wavelength range. Several experiments were conducted demonstrating frequency up-conversion of a single telecom-wavelength photon in a nonlinear crystal followed by detection in a Si APD [3] with high efficiency. A significant progress has been made to miniaturize and enhance the efficiency of second-order nonlinear optics using semiconductor materials [4, 5]. Nonetheless, in up-conversion schemes the underlying $\chi^{(2)}$ nonlinear interaction is relatively weak, and the overall detection process is a cascade of a non-resonant third-order process (as described by the time-dependent perturbation theory) followed by a resonant first-order process (Fig.1-a). Furthermore, the coherent non-resonant field interaction process requires complex dispersion compensation techniques - phasematching.

In contrast to insulating nonlinear crystals, semiconductors allow manipulation of free charge carriers, and thus more efficient resonant low-order nonlinear processes can be employed involving generation and recombination of charge carriers, which do not require phasematching. Moreover, semiconductor structures are easily fabricated using mature technology which allows for the realization of miniature devices in integrated photonic circuits. Semiconductor TPE [6] was recently proposed as a compact integrable source of entangled photons, essential for practical quantum information processing [7, 8, 9], three orders of magnitude more efficient [10] than the existing down-conversion schemes.

## II. IR SINGLE-PHOTON DETECTION BY TWO-PHOTON ABSORPTION

Here we propose a simple scheme for infrared single-photon direct detection in silicon (Si) APD assisted by a two-photon absorption (TPA) process [11] at room temperature, and develop the corresponding theoretical formalism. The energy of the detected photon at the signal frequency $\omega_s$ is lower than the bandgap and the energy difference is complemented by a pump field at $\omega_p$. The pump frequency $\omega_p$ is low enough to prevent TPA and three-photon absorption of the pump photons in Si (without the signal photon present), whereas the parasitic process of four-photon absorption is many orders of magnitude weaker and may be neglected. The proposed TPA based detection (Fig.1-b) is a second-order resonant process, and thus is more efficient than the fourth-order up-conversion scheme. Moreover, a significant virtue is that the conversion-detection is performed here in a single semiconductor device rather than a bulky scheme of nonlinear crystals and detectors.

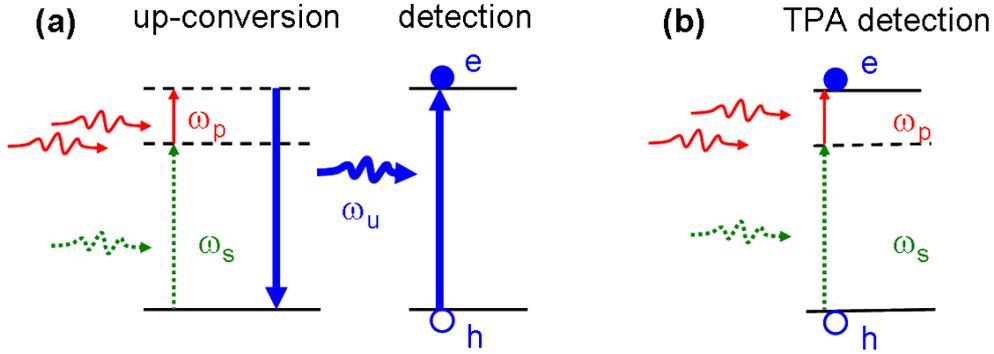

**Fig. 1. (a) detection via cascaded up-conversion and a detector; (b) detection via TPA. The solid horizontal lines represent electron energy levels and the dashed lines represent the virtual levels.**

The requirement that the pump frequency must be small enough to avoid the unwanted higher-order absorption dictates the TPA process in the proposed scheme to be very non-degenerate ($\omega_p \ll \omega_s$). Under such conditions existing theoretical models based on the perturbation theory fail, resulting in infrared divergences similar to those occurring in zero-frequency nonlinear optics calculations [12-13]. Non-perturbative solutions based on the dressed state approach yield infrared divergences as well when only a few elements of the series expansion are considered [14], which was shown to be analogous to the perturbative calculation result.

We develop a full non-perturbative dressed-state solution for the non-degenerate quantum TPA, where the infrared divergence is removed. In the presented model, the weak signal field is quantized as a single photon (to be detected), while the strong pump field is treated semi-classically. The conduction and the valence band electron wavefunctions are dressed by the strong pump field, while the weak signal field is assumed to have no effect on them. The system comprised of a semiconductor electron and an applied strong external field can be treated by Volkov function approach [15], where the transition occurs between the exact non-stationary wavefunctions, in contrast to the stationary approximations used in the perturbation theory. The current model is based on a single-particle approximation, which is reasonable for a room-temperature carrier-depleted bulk absorption region of the detector under reverse bias, while the minor many-body effect corrections are discussed in Sec. IV. The total energy of the quantum system neglecting the quadratic AC Stark effect [14] is:

$$H(\tau) = H_0 + H_{int}(\tau)$$
$$H_{int}(\tau) = -\frac{e\hbar}{m}\vec{A}_{dress}(\tau) \cdot \vec{k} \quad (1)$$

where $H_0$ is the energy of a particle associated with semiconductor energy band, $H_{int}$ the interaction energy between the standalone semiconductor electrons and a strong pump field, $\vec{A}_{dress}(\tau) = \hat{\varepsilon}_p A_p \cos(\omega_p t)$ is the dressing field vector potential, and $e$ and $m$ are the electron charge and effective mass respectively. The non-stationary Volkov wavefunction is:

$$|\psi\rangle = u(\vec{k},\vec{r})\exp\left\{i\vec{k}\cdot\vec{r} - \frac{i}{\hbar}\int_0^t H(\tau)d\tau\right\} \quad (2)$$

where $u(\vec{k},\vec{r})$ is the Bloch function and k is the electron crystal momentum.

The dressed state interaction can be visualized using Feynman diagrams. A crystal electron state is described as an interaction with static potential, similar to Bremsstrahlung depicted by a straight line (Fig. 2-a), while a dressed crystal electron state is represented by a zigzag line (Fig. 2-b) [16].

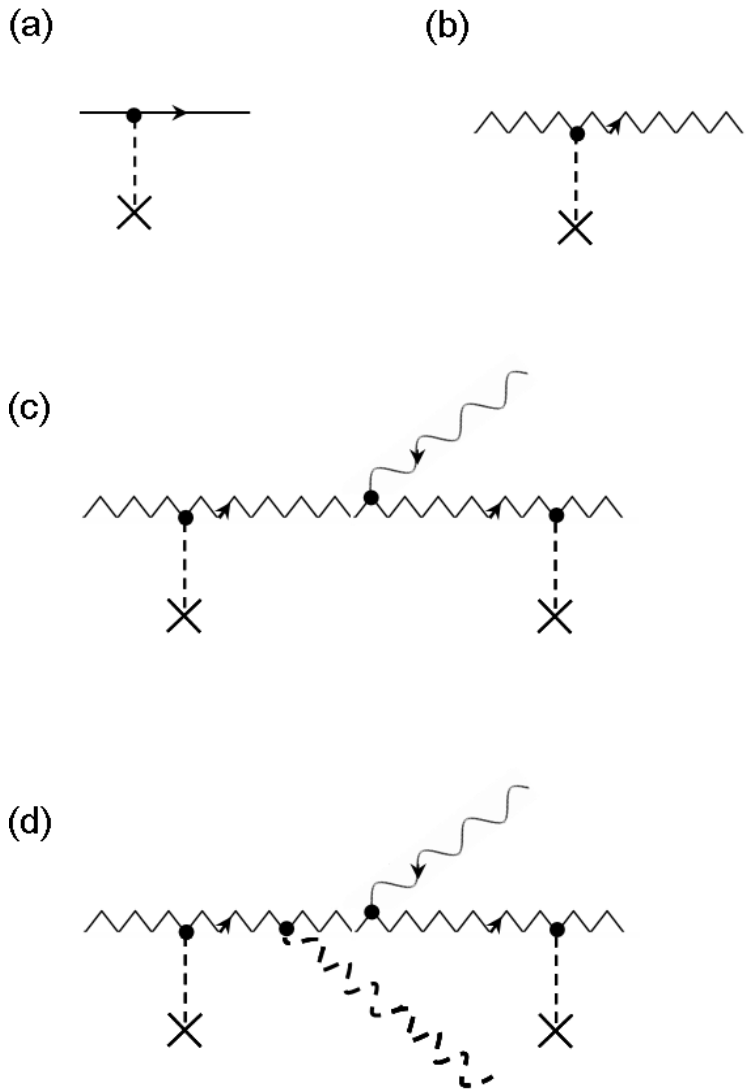

**Fig. 2. (a) Electron crystal state. The dashed line represents the electron-crystal interaction. (b) Electron dressed crystal state. (c) TPA in direct-band semiconductor- 1$^{st}$ order. The solid wavy line represents a photon (d) TPA in indirect-band semiconductor- 2$^{nd}$ order. The dashed wavy line represents a phonon**

In our model, for direct-band semiconductors, TPA is described by a first-order transition between dressed states accompanied by single photon absorption (Fig. 2-c), where the first-order interaction Hamiltonian is

$$H_{\text{int}}(\tau) = -\frac{e}{m_0}\sqrt{\frac{\hbar}{2\varepsilon_s V \omega_s}} \left( a_s^+ \hat{\varepsilon}_s e^{i\vec{k}\vec{r}} \cdot \hat{\vec{p}} + a_s \hat{\varepsilon}_s e^{-i\vec{k}\vec{r}} \cdot \hat{\vec{p}} \right) \quad (3)$$

with $\hat{\varepsilon}_s$ is the photon polarization, $\vec{k}$ is the photon wavevector, $\omega_s$ is the signal photon radial frequency, $\varepsilon_s$ is the material permittivity at $\omega_s$, $V$ is the field quantization volume, $\hat{\vec{p}}$ is the electron momentum operator, $m_0$ is the electron free-space mass, $\hat{a}_s$ and $\hat{a}_s^+$ are the field annihilation and creation operators.

## III. INDIRECT MATERIAL FORMALISM

In indirect bandgap semiconductors, such as Si, every optical transition is phonon-assisted [17]. Thus we describe TPA in indirect materials by a second-order perturbation applied to dressed states (Fig. 2-d). This dressed-state approach is much simpler than the more involved third order perturbation formalism. The interaction Hamiltonian, including the electron-phonon interaction $H_{e-phonon}$, is:

$$H_{\text{int}}(\tau) = -\frac{e}{m_0}\sqrt{\frac{\hbar}{2\varepsilon_s V \omega_s}} \left( a_s^+ \hat{\varepsilon}_s e^{i\vec{k}\vec{r}} \cdot \hat{\vec{p}} + a_s \hat{\varepsilon}_s e^{-i\vec{k}\vec{r}} \cdot \hat{\vec{p}} \right) \otimes H_{e-phonon} \quad (4)$$

The initial, final and intermediate states including the dressed electron wavefunctions, photon and the phonon Fock states are:

$$|\psi_i\rangle = u_v(\vec{k}_v, \vec{r})\exp\left\{i\vec{k}_v \cdot \vec{r} - \frac{i}{\hbar}\int_0^t H_v(\tau)d\tau\right\} \otimes |1_s\rangle \otimes |N_q\rangle$$

$$|\psi_f\rangle = u_c(\vec{k}_c + \vec{q}, \vec{r})\exp\left\{i(\vec{k}_c + \vec{q}) \cdot \vec{r} - \frac{i}{\hbar}\int_0^t H_c(\tau)d\tau\right\} \otimes |0_s\rangle \otimes |N_q \pm 1\rangle \quad (5)$$

$$|\psi_n\rangle = u_c(\vec{k}_c, \vec{r})\exp\left\{i\vec{k}_c \cdot \vec{r} - \frac{i}{\hbar}\int_0^t H_c(\tau)d\tau\right\} \otimes |0_s\rangle \otimes |N_q\rangle$$

Where the indices $v$ and $c$ designate the valence and the conduction bands, $\vec{q}$ is the phonon momentum. Momentum conservation in photon absorption process in indirect semiconductors requires emission or absorption of phonons, and therefore both cases $|N_q \pm 1\rangle$ are considered. The most significant contribution to the second-order electron-phonon interaction was shown to be made by considering the intermediate state as the higher energy minimum of the conduction band residing in the $\Gamma$ valley [18]. Second-order transition matrix element for a given electron k-state is then:

$$S = \left(\frac{1}{i\hbar}\right)^2 \sum_n \int_{-\infty}^{\infty} dt_1 \langle \psi_f | H_{int}(t_1) | \psi_n \rangle \int_{-\infty}^{t_1} \langle \psi_n | H_{int}(t_2) | \psi_i \rangle \tag{6}$$

yielding after the integration:

$$S = \frac{e}{m_0 \hbar^2} \sqrt{\frac{\hbar}{2\varepsilon_s V \omega_s}} \hat{\varepsilon}_s \cdot \vec{p}_{cv} \delta_{\vec{k}_c, \vec{k}_v} \int_{-\infty}^{\infty} dt_1 H^{\pm}_{e-phonon} e^{i(\omega_{cX} - \omega_{c\Gamma} \pm \omega_{phonon})t_1} \times$$
$$\int_{-\infty}^{t_1} dt_2 \cdot e^{i\omega_{c\Gamma} t_2} \left[ e^{\left(-\frac{ie}{m_c} \vec{k} \cdot \hat{\varepsilon}_p \frac{A_p}{\omega_p} \sin(\omega_p t_2)\right)} e^{(-i\omega_s t_2)} e^{\left(\frac{ie}{m_v} \vec{k} \cdot \hat{\varepsilon}_p \frac{A_p}{\omega_p} \sin(\omega_p t_2)\right)} \right] \tag{7}$$

where $\hbar\omega_{cX}$ and $\hbar\omega_{c\Gamma}$ are the conduction band X and $\Gamma$ valey energies respectively, $\vec{p}_{cv}$ is the inter-band transition matrix element, $H^{\pm}_{e-phonon}$ is the electron-phonon inter-valley phonon interaction matrix element, $m_c$ and $m_v$ are the conduction and the valence band effective masses. Defining a dimensionless parameter:

$$\eta_p = -\frac{eA_p}{m_{cv}\omega_p} \vec{k} \cdot \hat{\varepsilon}_p \tag{8}$$

with the reduced mass $m_{cv} = (1/m_c - 1/m_v)^{-1}$, considering a negative-charge negative-mass hole, the transition matrix element is:

$$S = \frac{e}{m_0 \hbar^2} \sqrt{\frac{\hbar}{2\varepsilon V \omega_s}} \hat{\varepsilon}_s \cdot \vec{p}_{cv} \delta_{\vec{k}_c, \vec{k}_v} \int_{-\infty}^{\infty} dt_1 H^{\pm}_{e-phonon} e^{i(\omega_{cX} - \omega_{c\Gamma} \pm \omega_{phonon})t_1} \times$$
$$\int_{-\infty}^{t_1} dt_2 \cdot e^{i\omega_{c\Gamma} t_2} \left( \sum_{n=-\infty}^{+\infty} J_n(\eta_p) e^{in\omega_p t_2} \right) e^{-i\omega_s t_2} \tag{9}$$

Where $J_n$ is the n-th order Bessel function of the first kind. From the energy conservation, the matrix element calculation results in:

$$S = -\frac{ie}{m_0\hbar^2}\sqrt{\frac{\hbar}{2\varepsilon V \omega_s}} \hat{\varepsilon}_s \cdot \vec{p}_{cv} \delta_{\vec{k}_c,\vec{k}_v} H^{\pm}_{e-phonon} J_{-1}(\eta_p) \frac{1}{\omega_{cv\Gamma} - \omega_p - \omega_s} 2\pi \delta(\omega_{cvX} - \omega_p - \omega_s \pm \omega_{phonon}) \quad (10)$$

Thus the absorption rate is [17]:

$$W = \frac{e^2 D_{ij}^2 (m_c m_v)^{3/2} |\hat{\varepsilon}_s \cdot \vec{p}_{cv}|^2}{16\pi^2 \hbar^5 \rho \omega_{phonon} \varepsilon \omega_s m_0^2 (\hbar \omega_{cv\Gamma} - \hbar(\omega_s + \omega_p))^2} \left[ \begin{array}{l} |J_1(\eta_p^+)|^2 N_q (\hbar(\omega_s + \omega_p) - \hbar\omega_{cvX} + \hbar\omega_{phonon})^2 + \\ |J_1(\eta_p^-)|^2 (N_q + 1)(\hbar(\omega_s + \omega_p) - \hbar\omega_{cvX} - \hbar\omega_{phonon})^2 \end{array} \right] \quad (11)$$

where $D_{ij}$ is the material deformation potential, $\rho$ is the material mass density, $\eta_p^+$ and $\eta_p^-$ correspond to phonon emission and absorption respectively and $N_q$ is the temperature-dependent phonon population.

## IV. DISCUSSION AND RESULTS

The model developed here treats the two-photon transition in direct and indirect semiconductors in the single-particle approximation. In general, a complete many-body approach introduces corrections to the single-particle formalism contributed by the excitonic effects [19] and carrier-carrier scattering processes affecting the exact estimation of absorption and gain coefficients in semiconductors including bandgap renormalization and Coulomb enhancement [20-21]. However, bandgap renormalization and Coulomb enhancement become insignificant in the depletion region of the detector, whereas the exciton-related absorption peaks at room-temperature practically disappear, only slightly modifying the absorption curve edge [17]. Moreover, in typical avalanche photodiode strong reverse bias, the carrier pairs generated in photon absorption are immediately swept apart by the strong filed, practically eliminating any electron-hole interaction [22]. For the specific material under consideration - Si, the Mott-Wannier exiton binding energy is about 11meV [23], smaller than the room-temperature thermal carrier energy ~26meV and much smaller than the carrier potential energy in the strong external field (~$10^6$ V/cm) [22] when an avalanche photodiode is considered. Furthermore, the potential TPA-based detector should be designed to operate at carrier energies above the band edges in order to maximize the absorption efficiency. Thus the exciton-induced band edge modification becomes much less important for the discussed application.

In order to demonstrate the practical feasibility of TPA-based infrared single-photon

APD detectors, Si-based waveguide APD [24] efficiency was calculated using the described model for practical device dimensions (~1mm length).

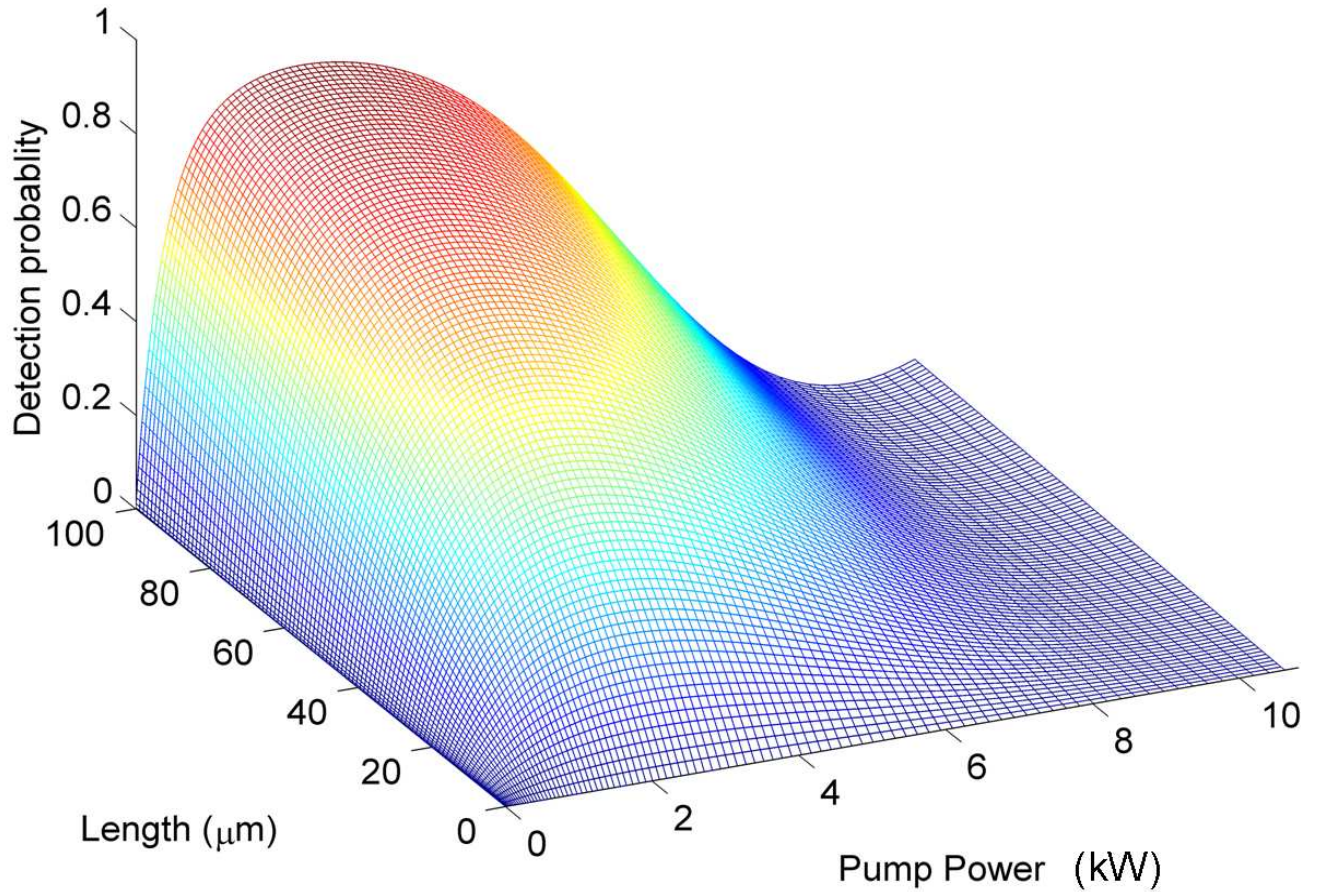

**Fig. 3. The probability of photon absorption as a function of the pump intensity and device length**

According to the calculations, the detection efficiency is increased (Fig. 3) with the device length and the increasing pump power until a certain high-field regime, after which the detection probability decreases. This effect can be attributed to the dressing-induced energy level detuning, which affects the photon absorption probability. Hence a certain optimal pump power can be calculated for a specific device. The calculated optimal pump power required for the device described here is around 1kW, which is ~1 order of magnitude lower than typical peak pump power used in up-conversion experiments [1] indicating the higher efficiency of the proposed TPA scheme.

## V. CONCLUSION

In conclusion, we have proposed an efficient nonlinear process of TPA in Si, which does not require phasematching, as the underlying mechanism for low dark-count room-temperature fast infrared photon counters. A quantum non-perturbative model is developed allowing proper non-diverging rate calculations and practical device optimization. Small dimensions and simple design of the theoretically demonstrated infrared photon-counters may enable integrated high-efficiency single-photon detection in the infrared for applications in astronomy, laser detection, and for quantum cryptography over telecommunication optical fibers.